\documentclass[10pt,sigconf,nonacm]{acmart}

\usepackage[all]{nowidow}
\usepackage{xcolor}
\usepackage{subcaption}
\usepackage{hyperref}
\usepackage{acronym}
\usepackage{csquotes}

\usepackage{siunitx}
\DeclareSIUnit[]{\flop}{\text{FLOP}}
\newcommand{\microvm}{\textmu{}VM}
\newcommand{\microvms}{\textmu{}VMs}

\usepackage[capitalise,noabbrev]{cleveref}
\crefformat{section}{\S#2#1#3} % see manual of cleveref, section 8.2.1
\crefformat{subsection}{\S#2#1#3}
\crefformat{subsubsection}{\S#2#1#3}
\crefrangeformat{section}{\S#3#1#4 to~\S#5#2#6}
\crefrangeformat{subsection}{\S#3#1#4 to~\S#5#2#6}
\crefrangeformat{subsubsection}{\S#3#1#4 to~\S#5#2#6}

\begin{document}

\author{Natalie Carl}
\authornote{Equal contribution.}
\affiliation{%
    \institution{Technische Universit\"at Berlin}
    \city{Berlin}
    \country{Germany}}
\email{nc@3s.tu-berlin.de}
\orcid{0009-0000-5991-9255}

\author{Tobias Pfandzelter}
\authornotemark[1]
\authornote{Part of this work was completed while Tobias Pfandzelter was at Technische Universität Berlin.}
\affiliation{%
    \institution{Unaffiliated}
    \city{Berlin}
    \country{Germany}}
\email{acm@pfandzelter.com}
\orcid{0000-0002-7868-8613}

\author{David Bermbach}
\affiliation{%
    \institution{Technische Universit\"at Berlin}
    \city{Berlin}
    \country{Germany}}
\email{db@3s.tu-berlin.de}
\orcid{0000-0002-7524-3256}

\title{Towards Energy-Efficient Serverless Computing with Hardware Isolation}

\begin{abstract}
    Serverless computing provides just-in-time infrastructure provisioning with rapid elasticity and a finely-grained pricing model.
    As full control of resource allocation is in the hands of the cloud provider and applications only consume resources when they actually perform work, we believe that serverless computing is uniquely positioned to maximize energy efficiency.

    However, the focus of current serverless platforms is to run hundreds or thousands of serverless functions from different tenants on traditional server hardware, requiring expensive software isolation mechanisms and a high degree of overprovisioning, i.e., idle servers, to anticipate load spikes.
    With shared caches, high clock frequencies, and many-core architectures, servers today are optimized for large, singular workloads but \emph{not} to run thousands of isolated functions.

    We propose rethinking the serverless hardware architecture to align it with the requirements of serverless software.
    Specifically, we propose using hardware isolation with individual processors per function instead of software isolation resulting in a serverless hardware stack that consumes energy only when an application actually performs work.
    In preliminary evaluation with real hardware and a typical serverless workload we find that this could reduce energy consumption overheads by \qty{90.63}{\percent} or an average \qty{70.8}{\mega\watt}.
\end{abstract}

\maketitle

\section{Introduction}
\label{sec:introduction}

Serverless or Function-as-a-Service (FaaS)\footnote{While there are competing definitions for these terms, we use them interchangeably in this paper.} allows developers to rent cloud infrastructure with millisecond-granularity and rapid elasticity~\cite{jonas2019cloud,hendrickson2016serverless,hellerstein2018serverless}.
Applications are composed as collections of stateless functions that are invoked in response to external and internal events, e.g., HTTP requests, messages, or as part of function workflows~\cite{jonas2019cloud,eismann2021serverless}.
Cloud serverless platforms, such as \emph{AWS Lambda}~\cite{lambda} or Google Cloud Run Functions~\cite{gcf}, schedule isolated worker environments on their datacenter infrastructure to quickly execute these small cloud functions.

One of the key challenges in the efficient execution of cloud serverless functions is that of isolation~\cite{jonas2019cloud}.
Individual function executions usually only have access to few compute resources, such as a single or even a fraction of a CPU core.
Cloud providers are thus interested in running tens, hundreds, or thousands of functions on the same physical server, without sacrificing isolation between tenants~\cite{wang2021faasnet,cvetkovic2024dirigent}.
Research has shown that full-blown virtual machines have too large of a footprint to enable this efficiently, while most process-level containerization does not provide the required levels of isolation, which has spurred innovations such as \microvms{} and hardened containers~\cite{agache2020firecracker,gvisor2018}.

Taking a step back from these advancements, we notice an interesting trend:
On the one hand, commercial cloud actors and serverless research have invested heavily in efficiently fragmenting servers into small, isolated environments for function execution.
On the other hand, in an exact opposite trajectory, those researching, developing, and building servers and datacenter infrastructure aim to transparently scale infrastructure up and out, manifesting in multicore systems with shared memory and caches or high-speed, high-capacity interconnects between nodes with direct memory access~\cite{verma2015large,pemberton2019serverless,baker2000cluster}.
Put simply, current serverless platform architectures aim to make one large computer --- the datacenter --- look like many miniscule computers, while current infrastructure aims to make many small computers --- servers and individual CPU cores --- appear as one large computer.

\begin{figure}
    \centering
    \begin{subfigure}{0.49\linewidth}
        \centering
        \includegraphics[width=\linewidth]{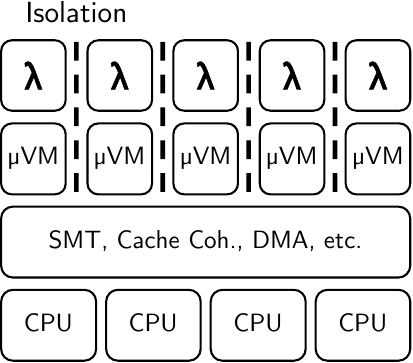}
        \caption{Traditional isolation}
        \label{fig:lambda:old}
    \end{subfigure}
    \hfill
    \begin{subfigure}{0.49\linewidth}
        \centering
        \includegraphics[width=\linewidth]{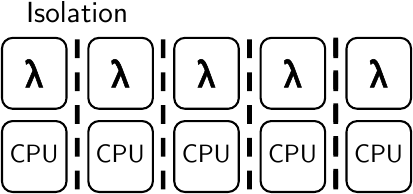}
        \caption{Hardware-level isolation}
        \label{fig:lambda:new}
    \end{subfigure}
    \caption{
        Traditional cloud serverless platforms use software-level isolation, e.g., using \microvms{}, on top of shared hardware built to make many cores or servers appear as a singular computer (\cref{fig:lambda:old}).
        We propose running functions in a one-to-one relationship on isolated, much smaller computers to decrease energy overheads (\cref{fig:lambda:new}).
    }
    \label{fig:lambda}
\end{figure}

Of course, these technologies exist for a reason, enabling efficient deployment for a majority of applications.
However, we argue that research has not investigated hardware architectures for serverless computing sufficiently.
We believe that especially energy efficiency and density of serverless platforms can be further improved by reimagining serverless hardware.
Our vision is simple, as shown in \cref{fig:lambda}:
Instead of software-level isolation on top of abstract many-CPU infrastructure, we propose executing cloud functions directly on isolated computers, eliminating efficiency overheads for both isolation and server-scaling technologies.
Further, we argue that a direct relationship between function execution and the underlying hardware allows more efficient execution.

\section{Servers vs.~Serverless}
\label{sec:mismatch}

Before elaborating on our vision, we give an overview of the three main inefficiencies we identify for traditional serverless platform architectures.

\subsection{Server infrastructure is not built for many small, independent processes}
\label{sec:mismatch:servers}

We can imagine the execution of serverless functions as a collection of small (access to fractions of a regular CPU core's cycles), short-lived (on the order of seconds), independent (no direct interaction between them) processes.
It is not difficult to see how trends in server architecture have targeted diametrically opposite use cases over the past few years.

On the level of an individual CPU core, the decline of Dennard scaling and the end of Moore's law have reduced the year-over-year single-core performance increases of CPUs~\cite{esmaeilzadeh2011dark}.
Still, modern CPU cores have achieved such high performance that fractions of core cycles are sufficient for many serverless functions; e.g., the default for AWS Lambda is one twelfth of a virtual CPU, which itself is only half a physical CPU core but leverages simultaneous multithreading~\cite{schirmer2024fusionizepp,ec2vcpuoptions}.
As a result, frequent preemption is necessary to support full provisioning, which has been shown to result in suboptimal performance and, hence, performance efficiency~\cite{fu2022sfs,isstaif2023latency,zhao2024serverless}.

One step up from here, servers usually include tens or hundreds of CPU cores on one or multiple physical CPUs.
Given the physical limits of CPU core performance, having multiple cores is of course highly beneficial for applications that can distribute their work across multiple threads.
Hardware and operating system features such as shared memory and caches can further increase the efficiency of such applications by lowering communication and synchronization costs between cores.
For most serverless functions, which may access at most two or three full physical CPU cores (AWS Lambda functions are limited to six vCPUs~\cite{lambdamaxvcpu}), such synchronization mechanisms are not only irrelevant but may decrease system efficiency by taking up die space.

Similar technologies at the level of clusters of servers, e.g., remote direct memory access (RDMA) and other high-speed interconnects, may be similarly ill-suited for serverless workloads.
In turn, we find it unlikely that they are widely in use in serverless infrastructure (although their use has been investigated in research~\cite{wei2023mitosis,copik2023rfaas}).

\subsection{Software-level isolation requires indirection}
\label{sec:mismatch:isolation}

Executing different small serverless functions on a single, large node requires isolation mechanisms, such as the \microvms{} and hardened containers that we have alluded to.
There is a trade-off between the level of isolation such sandboxing mechanisms afford and their overhead.
On the one hand, \microvms{} provide a high level of isolation but require more time and energy to start as well as indirection for critical tasks, such as system calls~\cite{young2019true,sharma2023challenges}.
On the other hand, containers have little overhead but provide insufficient isolation~\cite{moebius2024unikernel}.
Some researchers have proposed reducing isolation for trusted workloads, e.g., concurrently running invocations for the same function in a single sandbox~\cite{jia2021nightcore,akkus2018sand}.

Commercial serverless platforms decidedly fall on the end of the spectrum that values isolation over performance, explaining the development of \emph{Firecracker} \microvms{} for AWS Lambda~\cite{agache2020firecracker} and \emph{gVisor} hardened containers for Google Cloud Run Functions~\cite{gvisor2018}.
To decrease the impact of expensive sandbox creation, most serverless platforms, e.g., AWS Lambda, follow a simple protocol for function invocations~\cite{brooker2023demand}:
When a request for a function comes in, it is forwarded to an idle worker for that specific function (\emph{warm start}).
If no idle worker is available, a new worker is allocated (\emph{cold start}).
The specifics of worker allocation differ between platforms, with AWS Lambda using ready, empty \microvms{} that pull in function images from a central location.
Crucially, once a worker has completed the function invocation, it remains idle and available for future invocations of that same function for some amount of time, e.g., AWS Lambda evicts half the number of idle workers for a function every \qty{380}{\second}~\cite{copik2021sebs}, whereas OpenFaaS lets operators configure a static idle timeout, e.g., \qty{15}{\minute}~\cite{openfaastimeout}.
During that time, subsequent invocations of that function benefit from faster warm starts, yet the infrastructure of that worker is blocked for other functions, requiring significant energy~\cite{roy2024hidden}.

\subsection{Granularity mismatch between large servers and small functions}
\label{sec:mismatch:granularity}

Even with overcommitment, servers that host idle workers are unlikely to ever be fully utilized, simply as a result of the elasticity of serverless workloads.
A server that could host, say, one hundred function workers but currently executes only a single function requires more energy per function invocation than a fully utilized server.
Further, to provide rapid elasticity, a number of empty servers must be kept ready to host new workloads with little notice~\cite{agache2020firecracker}.
At the same time, hotspots, where servers are fully utilized, also lead to suboptimal power states.
This mismatch in granularity between large, slow-to-start servers and small, rapidly elastic functions with unpredictable workload patterns makes scheduling non-trivial.
The resulting nonlinear energy consumption of serverless computing is far from efficient at most times~\cite{rehman2024faasmeter,sharma2024accountable}.

\section{Hardware Isolation for Serverless Computing}
\label{sec:proposal}

\begin{figure}
    \centering
    \includegraphics[width=0.8\linewidth]{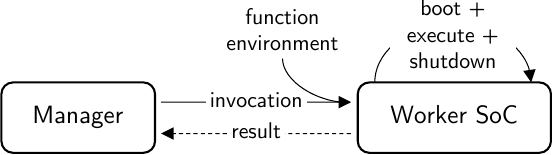}
    \caption{The lifecycle of our proposed approach: A manager sends the function invocation (including possible input parameters) to an available worker SoC. This worker boots from an environment for this particular function and executes that function. If necessary, the result of the execution can be sent back to the manager. As soon as the function has ended, the worker shuts down.}
    \label{fig:lifecycle}
\end{figure}

Our vision for serverless hardware infrastructure is simple:
Instead of a fleet of small virtual machines, i.e., \microvms{}, on cluster hardware, we propose leveraging a fleet of small physical machines.
These are not necessarily individual rack-mounted servers but can also be many small, independent system-on-a-chips (SoCs) in a shared enclosure or even on the same die.
When executing a function, the SoC boots into the function's execution environment, e.g., a lightweight Linux or even unikernel with the function deployment package, performs the execution, and then stops and shuts down.
We illustrate this lifecycle in \cref{fig:lifecycle}.

Our vision is simple, yet it directly addresses the inefficiencies we have outlined in \cref{sec:mismatch}.
First, leveraging small SoCs allows using more simple chip designs, such as energy-efficient mobile chips.
Instead of allocating, e.g., an eight of the CPU cycles of a high-performance server CPU core, we may use a smaller, single-core chip with an eight of the performance, possibly with a better efficiency profile.
Running a single process on this chip also alleviates overheads by preemption or shared caches or memory.

Second, running functions directly on isolated hardware obviates the need for software-level isolation mechanisms, removing the energy impact of starting \microvms{} or containers.
What remains are simple deployment packages for functions, akin to container images.
By booting into these images and shutting down the system after execution, idle times of the hardware, i.e., those not spent executing a function, are reduced to a minimum.

Third, this also results in a total energy consumption that is linear with the rate and duration of function execution, which is much more fitting for the rapidly elastic nature of serverless workloads.

\section{Preliminary Evaluation}
\label{sec:evaluation}

To illustrate the efficacy of our approach, we analyze the pattern of a typical serverless workload and extrapolate its energy consumption on commodity hardware.
We then present a preliminary prototype of our approach on a small single-board computer and measure its energy overhead, allowing us to calculate potential performance efficiency gains.
We make all artifacts used to produce these results and our collected data available.\footnote{\url{https://github.com/3s-rg/chipless}}

\subsection{Simulating elastic serverless workloads}
\label{sec:evaluation:simulation}

Serverless workloads are inherently elastic, with function invocations correlating directly with external API or web calls.
While there have been proposals for workload smoothing with delayed function invocations, e.g.,~\cite{schirmer2023profaastinate,sahraei2023xfaaS,segarra2024cold}, we assume here that a serverless function should be executed as soon as it is invoked~\cite{jonas2019cloud}, given the limited benefits of temporal workload shifting for sustainability in practice~\cite{sukprasert2024limitations}.

Based on the traditional serverless function execution mechanism outlined in \cref{sec:mismatch:isolation}, we simulate the resource usage of serverless infrastructure with a 24-hour subset of production workload traces from the 2023 Huawei internal serverless workload data set~\cite{joosen2023how}.
This data set contains invocations and execution durations for \num{200} functions at per-second granularity.
We assume that all functions have equal resource limits, which is simplified but would lead to more efficient resource allocation.
We specifically simulate the worker allocation necessary to handle this workload, measuring how many workers are idle, busy, or newly started for each one-second timestep in the simulation.
Here, we assume that workers remain idle and available for \qty{15}{\min} after executing a function and our scheduler preferring workers with lower idle time (further increasing allocation efficiency).
Note that a higher idle time will inevitability lead to more idle workers, while a lower idle time will increase the number of cold starts for the function.

\begin{figure*}
    \centering
    \includegraphics[width=\linewidth]{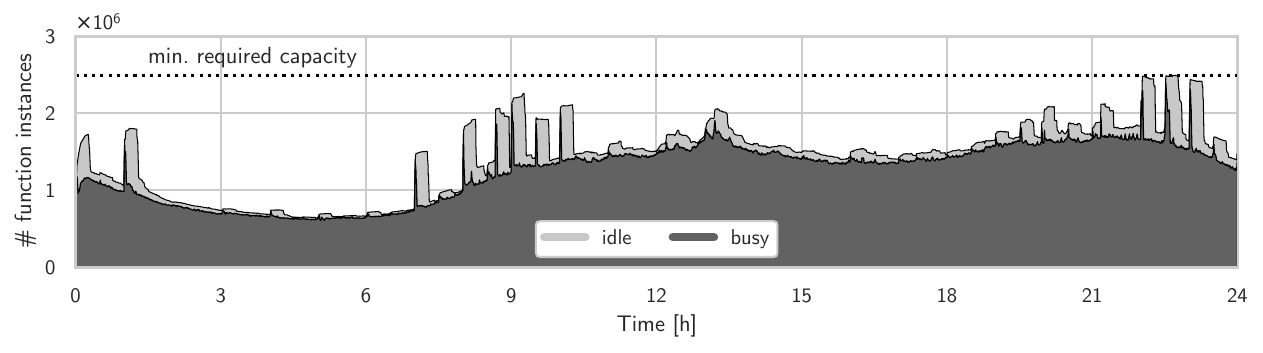}
    \caption{Number of function instances, or workers, across all 200 functions in the workload trace over the course of 24$\,$h. Periodic workload spikes leave unused idle workers that block resources. The dashed line indicates the minimum capacity required in the serverless infrastructure to handle all events at the same service level.}
    \label{fig:simulation}
\end{figure*}

The results in \cref{fig:simulation} reflect the typical variation of serverless workloads throughout a day~\cite{sahraei2023xfaaS,schirmer2023nightshift,joosen2023how}.
Load spikes periodically lead to higher capacity utilization, after which workers created to handle these additional requests remain idle and unused.
In addition to resources blocked by idle workers waiting for their next invocation, the maximum number of workers throughout the day also reflects the minimum required capacity of serverless infrastructure to ensure that each request can be handled with sufficient resources.
For the particular workload trace we consider in our simulation, this minimum capacity is approximately \num{2.49} million concurrent workers.
To put this into perspective, our workload trace has an average of \num{49386.85} requests per second, while AWS Lambda reportedly serves on the order of \num{4} million requests per second across all availability zones~\cite{lambdainvocations}.
The simulation results clearly show how idle function workers block available resources.
At the same time, large amounts of empty capacity may also have an impact on energy efficiency if they cannot be turned off or used for other productive work~\cite{fbautoscale}.

\subsection{Energy use of isolation approaches}
\label{sec:evaluation:energy}

To translate the results of our simulation into energy requirements, we compare the traditional \microvm{} serverless isolation approach with our proposal of running each function on dedicated hardware.
To that end, we use two devices, each fitted with a socket-level power meter.
First, our server hardware is a rack-mounted 1U server with two 12-core Intel Xeon \num{4310} CPUs (\num{48} virtual cores with simultaneous multithreading), \qty{64}{\gibi\byte} of memory, and two SAS SSDs.
Second, we use a SinoVoip Banana Pi M2 Zero single-board computer with a quad-core Allwinner H3 ARM CPU, \qty{512}{\mebi\byte} of memory, and \textmu{}SD storage.
Both devices run in room-temperature environments.

The server draws \qty{120}{\watt} of power at idle, which translates to \qty{2.5}{\watt} per virtual core, and up to \qty{330}{\watt} under full load.
The single-board computer draws only \qty{0.6}{\watt} at idle and up to \qty{3.6}{\watt} under full load.
In preliminary LINPACK benchmarks, however, we found the server to be more than five times as power efficient as the single-board computer, at \qty{2.58}{\giga\flop\per\watt} and \qty{0.45}{\giga\flop\per\watt}, respectively.
Presumably, this is a result of the old design of the Allwinner H3 chip, which was first released in 2014.
For the purposes of this paper, this means that any efficiency results we achieve with our vision on this particular hardware are not a result of more efficient chips but rather \emph{despite} inefficient chips.
Nevertheless, we use this hardware for our experiments for availability and compatibility reasons, yet we acknowledge that exact results may be slightly different on other hardware.

We measure the energy requirement of \microvms{} by booting between one and \num{96} (twice the number of available virtual cores) single-vCPU Firecracker \microvms{} on our server.
In our measurement, we include preparation of network devices and \texttt{cgroup} creation for the \microvm{}, boot time of a small Firecracker Linux kernel, and initialization time of a stripped-down Alpine Linux distribution until the virtual machine can perform a network callback to our measurement host.
We repeat each experiment three times.

\begin{figure}
    \centering
    \includegraphics[width=\linewidth]{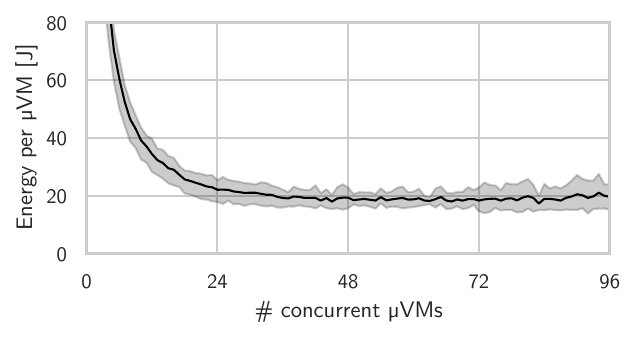}
    \caption{Energy requirements per \microvm{} start are highest for a single \microvm{} (335.81$\,$J, outside the area of the graph), as most server cores remain idle. Starting \microvms{} is most efficient when booting 24 to 48 \microvms{} concurrently, or between one and two per physical core. Still, energy use per \microvm{} start is high at a minimum 17.98$\,$J.}
    \label{fig:firecracker}
\end{figure}

As shown in \cref{fig:firecracker}, energy use is highest when booting only a single \microvm{}, at a mean \qty{335.81}{\joule} (outside the area of the graph).
This is expected, as booting the single \microvm{} leaves all but one CPU core idle.
This single \microvm{} boot takes a mean \qty{2.47}{\second}, which includes not just the relatively fast kernel boot time (Agache et al.~\cite{agache2020firecracker} find Firecracker \microvms{} to take on the order of \qty{200}{\milli\second} to boot) but also configuring the Firecracker jailer sandbox, including copying kernel and boot volumes into the sandbox, configuring the \microvm{} through Firecracker's HTTP interface, configuring network interfaces on the host and inside the \microvm{}, and running a lightweight \texttt{init} script.
Conversely, per-\microvm{} energy consumption is lowest when booting \num{48} \microvms{}, one for each virtual CPU core of the server.
Here, mean energy consumption per \microvm{} is only \qty{17.98}{\joule}, which is in line with the results of Sharma~\cite{sharma2023challenges}.
Idle power draw remains at \qty{120}{\watt} for the server, even with \num{48} idle \microvms{}, i.e., \qty{2.5}{\watt} per \microvm{}.

As a preliminary prototype of our proposal, we measure the power consumption of a boot of our single-board computer into a reduced Linux environment, from enabling a relay (powering on the system) to receiving a network callback at the measurement computer.
This reduced Linux environment would be able to perform the same function execution as a \microvm{} in the previous experiment, except for differences in microarchitecture between the two systems.
We control our SoC with a software-controlled relay.

\begin{figure}
    \centering
    \includegraphics[width=\linewidth]{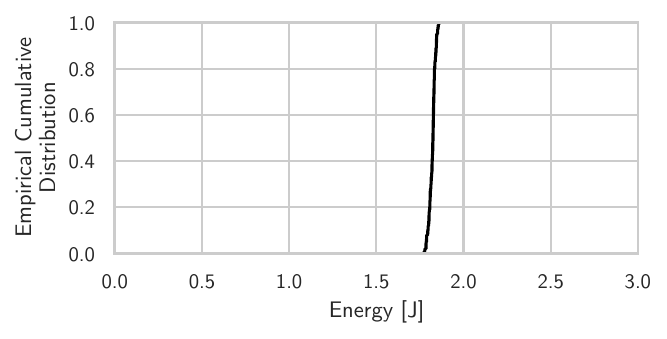}
    \caption{100 repetitions of booting our SoC into a lightweight Linux environment show stable energy use of a mean 1.83$\,$J per boot. This is an order of magnitude smaller than booting a \microvm{} on our server.}
    \label{fig:banana-chipless}
\end{figure}

We show the energy measurements of \num{100} repetitions of booting the SoC in \cref{fig:banana-chipless}.
At a mean \qty{1.83}{\joule} per boot, energy use is an order of magnitude smaller than booting \microvms{} on our server in the most efficient conditions, despite the age and inefficiency of the SoC we use in this prototype.
This is a significant reduction in energy use for function sandboxing.

Boot times are longer than for \microvms{} at a mean \qty{3.16}{\second}, which is an infeasible cold start time for production workloads.
Interestingly, the actual Linux kernel boot time is consistently on the order of only \qty{77}{\milli\second}, with most time spent on the SoC bootloader.
In this particular case, we suspect two components to slow boot significantly, although we were unable to trace their impacts precisely:
First, the open-source \emph{U-Boot} bootloader~\cite{uboot} for our SoC uses a multi-stage boot process that involves entering a full boot environment, including debugging capabilities.
We have reason to believe that bypassing this environment, e.g., using U-Boot's \emph{Falcon} boot mode~\cite{falcon}, can reduce this overhead.
Second, booting from \textmu{}SD storage is comparatively slow.
In summary, we believe that this cold start overhead can be reduced significantly with further research into system architecture (see also \cref{sec:outlook}).
Still, we note that the goal of our proposal is not to make serverless function execution faster -- ideally, we can increase serverless energy efficiency while maintaining cold start and execution performance.

\subsection{Extrapolating serverless energy use}
\label{sec:evaluation:extrapolation}

\begin{figure}
    \centering
    \includegraphics[width=\linewidth]{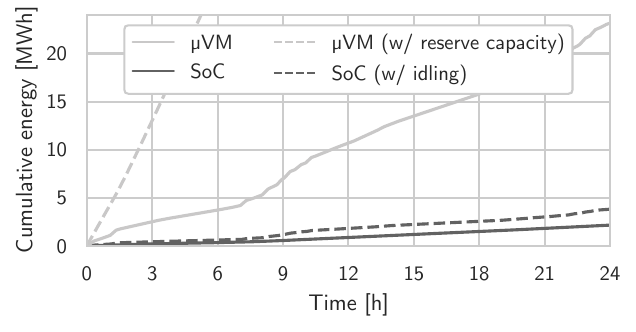}
    \caption{Extrapolating excess energy use for our workload over the course of 24$\,$h shows that our approach can significantly increase energy efficiency for serverless infrastructure. While \microvms{} lead to 22.32$\,$MW$\,$h excess energy, using SoCs reduces this by 90.28$\,$\%, to 2.17$\,$MW$\,$h.}
    \label{fig:extrapolation}
\end{figure}

Based on our simulation of production workload traces in \cref{sec:evaluation:simulation} and energy measurements in \cref{sec:evaluation:energy}, we can now extrapolate \emph{wasted} energy in serverless infrastructure, i.e., energy required to start worker sandboxes and to cache idle workers.
We assume that a worker executing a function, whether in a \microvm{} or on an SoC, uses all its energy productively, i.e., at full efficiency.
We consider the cumulative excess energy for our workload over its span of \qty{24}{\hour}, comparing \microvms{}; \microvm{} workers plus idle energy of reserve worker capacity; SoCs with function execution as outlined in \cref{sec:proposal}; and SoCs that are also allowed to become idle, i.e., are not shut down directly after function execution but instead allowed to execute future invocations for the same function.
For \microvms{}, we assume the measured \qty{17.98}{\joule} per worker to start a function instance and \qty{2.5}{\watt} power draw while a worker is idle.
When also considering energy use of reserve capacity, we assume \qty{2.5}{\watt} for all capacity that is not busy executing a function.
For SoCs, we calculate only \qty{1.83}{\joule} per worker start, i.e., per incoming request, and no idle power draw.
To assess the impact of removing worker idling, we also extrapolate the energy use of our proposal if SoCs workers were allowed to remain idle, at a \qty{0.6}{\watt} idle power draw (measured using the same SoC as in \cref{sec:evaluation:energy}).

The results in \cref{fig:extrapolation} clearly illustrate the potential of our vision.
Over the course of this \qty{24}{\hour} workload, using \microvms{} for worker isolation leads to \qty{23.15}{\mega\watt\hour} excess energy use, whereas our SoC approach would only require \qty{2.17}{\mega\watt\hour} (a reduction of \qty{90.63}{\percent}).
On average, this is equal to a reduction of \qty{874.16}{\kilo\watt} in power draw.
Recall that AWS Lambda processes more than eighty times as many requests as in our workload trace.
With linear extrapolation, power draw could be reduced by roughly \qty{70.8}{\mega\watt} at this scale, before accounting for datacenter power usage efficiency~\cite{sharma2024jevons}.
This is roughly equivalent to the maximum capacity of \num{20} average U.S.~wind turbines~\cite{windturbines}.

This assumes that reserve capacity in the serverless infrastructure does not draw any energy, e.g., by using it for productive work or turning it off and on with second-granularity.
When accounting for the energy use of this reserve capacity, we calculate a \qty{86.86}{\mega\watt\hour} upper bound for the wasted energy produced by serverless infrastructure for our workload.
Of course, these numbers are also likely to be higher for larger workloads.

Interestingly, letting SoC workers idle also reduces energy efficiency, increasing excess energy use only to \qty{3.82}{\mega\watt\hour}.
This is not unexpected, as the idle energy use of our SoC exceeds its boot power consumption after only a few seconds ($\frac{\qty{1.83}{\joule}}{\qty{0.6}{\watt}} = \qty{3.05}{\second}$), far quicker than the \qty{15}{\minute} timeout for workers we use in our simulation.

\section{Outlook}
\label{sec:outlook}

Our results are preliminary but indicate a significant opportunity for the reduction of energy use in serverless cloud computing.
The approach we have outlined in this paper can make this possible, but our vision is merely the starting point and a number of research questions are still open for future investigation.

\paragraph*{Are SoCs fit for the datacenter?}

As is the main point of our argument, using SoCs in a datacenter is a significant departure of well-known datacenter infrastructure.
The SoC we use in our preliminary evaluation is merely a development board, whereas production deployment in a datacenter will require further development of boards of SoCs that are compatible with the realities of large-capacity cooling and power supply.
Interestingly, Zhang et al.~\cite{zhang2024more,zhang2024high,xu2022position} report extensively on the development, evaluation, and commercialization of more than \num{10000} SoC clusters for edge computing and cloud gaming applications.
Their cluster servers comprise \num{60} Qualcomm Snapdragon 865 SoCs in a 2U rack, which results in high capital expenditure compared to traditional CPU-only server hardware yet at significantly higher energy efficiency.
Crucially, this indicates that building SoC servers is feasible.
Similar proposals to leverage SoC hardware for server infrastructure have been put forwarded in the past, yet without evaluation at this scale~\cite{janapa2010websearch,buesching2012droidcluster}.
Future work should focus on how SoCs can be best integrated into the datacenter.

\paragraph*{Can we further reduce the abstractions between functions and hardware?}

Our vision is to completely remove the abstractions between serverless functions and their underlying hardware to make serverless execution as efficient as possible.
In our preliminary experiments, however, we still rely on an operating system running between function executable and SoC, namely Linux.
In fact, with a single application, often even a single process running on our hardware, this may be an unnecessary layer of abstraction.
An interesting avenue for future work is the investigating of deployment mechanisms without operating systems, such as unikernels.
Unikernels have already been investigated for application in serverless platforms, but usually on top of hypervisors such as \microvms{}~\cite{moebius2024unikernel,parola2024sure}.
G{\'e}hberger and Kov{\'a}cs~\cite{gehberger2022cooling} have proposed leveraging unikernels for the elimination of idle instances and warm starts completely, similar to our approach.
With a smaller footprint and faster boot times, unikernels could thus further increase serverless efficiency, yet deploying them directly on hardware is a difficult endeavor that requires further research.

\paragraph*{Will reducing serverless energy use also reduce serverless environmental impacts?}

Increasing the energy efficiency of serverless computing is certainly useful, but not necessarily intrinsically a desirable goal.
In the short term, our proposal could reduce operational expenditure for cloud providers by reducing the power draw of their serverless workloads.
Of course, this comes at the cost of capital expenditure for new types of servers, which is not unreasonable in the short term given the typical three-year lifespan of datacenter equipment~\cite{barroso2019datacenter,zhang2024more}.
In the long term, however, the environmental impact of our proposal is not as easily assessed as assuming that higher energy efficiency also leads to lower carbon emissions.
First, the impact of embodied carbon of specialized serverless hardware remains to be investigated.
One possible way to address this is the use of recycled mobile hardware, although hardware efficiency and integration efforts may make this infeasible~\cite{buesching2012droidcluster,switzer2023junkyard,shahrad2017decommissioned}.
Second, there is a correlation between the energy efficiency of cloud datacenters and their total energy consumption that requires further research~\cite{sharma2024jevons}.
Counterintuitively, increasing energy efficiency, thus lowering cost, could lead to higher adoption and thus higher total energy use.
Third, some cloud providers already claim carbon neutrality for their datacenters, while others are investigating building new nuclear power plants~\cite{nuclear}.
Still, concerns about noise pollution, grid strain, and water use remain~\cite{ngata2025cloud}.

\paragraph*{How much flexibility does serverless resource allocation require?}

By replacing \microvms{} and other container technologies with SoCs, we lose flexibility in serverless resource allocation.
Currently, customers of AWS Lambda can specify memory allocations for their functions with megabyte-granularity, directly impacting how long function execution takes and how much each invocation costs~\cite{lambda}.
With static SoCs, this allocation is fixed, at least to the few different sizes of SoCs that the cloud provider installs in their datacenter.
Choosing the right sizes for such SoCs is thus an interesting problem:
On the one hand, SoCs too large are a wasted investment for smaller functions.
On the other hand, SoCs that are too small are unable to support larger functions.
Further research on the true resource requirements of serverless functions is necessary, with a specific focus on whether tenants really need megabyte-granularity for resource allocation.
Interestingly, research on tuning serverless functions has assumed that tenants are mostly interested in achieving minimum execution latency or cost, which would still be possible with a more limited selection of SoC sizes~\cite{schirmer2024fusionizepp,eismann2021sizeless,cordingly2022function,powertuning}.

\section{Conclusion}
\label{sec:conclusion}

In this paper, we have outlined the mismatch between cloud serverless software and serverless hardware infrastructure.
While serverless platforms aim to create thousands of small, isolated execution environments for functions, traditional datacenter hardware aims to layer an abstraction of a singular execution environment on top of multicore machines.
We proposed breaking these abstractions with hardware-level isolation for serverless functions to increase the energy efficiency of cloud serverless computing.
Specifically, we imagined each function having access to a small, isolated computer, such as an SoC, for the duration of its execution.
This computer could be turned off when not executing a function, eliminating idle power draw.
In preliminary evaluation on real hardware and with real serverless traces, we found this to reduce excess energy use by \qty{90.63}{\percent}.

\begin{acks}
    We thank Leon P\"ollinger, Timo Oeltze, Gabriel Behrendt, Vincent Clermont, and Petros Kanellopoulos for performing a preliminary feasibility evaluation of our approach in the scope of a student project.
    Funded by the \grantsponsor{BMBF}{Bundesministerium für Bildung und Forschung (BMBF, German Federal Ministry of Education and Research)}{https://www.bmbf.de/bmbf/en} (\grantnum{BMBF}{16KISK183}) and
    \grantsponsor{DFG}{Deutsche Forschungsgemeinschaft (DFG, German Research Foundation)}{https://www.dfg.de/en/} (\grantnum{DFG}{495343202}).
\end{acks}

\balance

\bibliographystyle{ACM-Reference-Format}
\bibliography{bibliography.bib}

\end{document}